\documentclass[12pt,a4paper]{article}
\usepackage{amsmath}
\usepackage{amssymb}
\usepackage{amsthm}
\usepackage{float}
\usepackage{amsfonts}
\usepackage{graphicx}
\usepackage[]{cite}
\usepackage{float}
\usepackage{subfigure}
\usepackage{verbatim}
\usepackage[left=2cm,right=2cm,top=3cm,bottom=2.5cm]{geometry}
\usepackage[numbers]{natbib}
\usepackage[utf8]{inputenc}
\def\case#1/#2{\frac{#1}{#2}}

\def \D {\tilde{\nabla}}
\def\la {\langle}
\def\ra {\rangle}
\newcommand{\sfrac}[2]{{\textstyle{#1\over#2}}}
\def \ep {\varepsilon}
\def\tl{\tilde}
\def\rd {\displaystyle{\cdot}}
\def\ts {\textstyle}

\usepackage[usenames,dvipsnames,svgnames]{xcolor}
\usepackage[colorlinks=true,
      linkcolor=red,
      urlcolor=gray,
      citecolor=blue]{hyperref}

\def\myalign#1{%
  \def\trule{\noalign{\smallskip\hrule\medskip}}
  \def\nebc{\nearrow\bigcup}
  \def\sebc{\searrow\bigcup}
  \def\pminf{{}_{-\infty}|^{+\infty}}
  \let\Inf\infty
  \def\amp{&} 
  \vbox{\mathsurround0pt\openup1\jot
    \halign{%
      &$\displaystyle##\hfil\tabskip0pt$&\amp##\tabskip1em\crcr
      \noalign{\hrule height1pt\smallskip}#1\noalign{\smallskip\hrule height1pt}\crcr}}}

\begin{document}
\begin{center}
\textbf{Perturbations in the interacting vacuum}
\end{center}
\hfill\\
Albert Munyeshyaka$^{1}$,Joseph Ntahompagaze$^{2}$,Tom Mutabazi$^{1}$, Manasse.R  Mbonye$^{2,3,4}$, Abraham Ayirwanda$^{2}$,Fidele Twagirayezu $^{2}$ and Amare Abebe$^{5,6}$\\
\hfill\\ 
%$^{1}$Department of Physics, College of Science and Technology, University of Rwanda, Rwanda\;\;\; \; \hfill\\
$^{1}$Department of Physics, Mbarara University of Science and Technology, Mbarara, Uganda\;\;\; \; \;\hfill\\
$^{2}$Department of Physics, College of Science and Technology, University of Rwanda, Rwanda\;\;\; \; \;\hfill\\ 
$^{3}$ International Center for Theoretical Physics (ICTP)-East African Institute for Fundamental Research, University of Rwanda, Kigali, Rwanda
\;\;\; \; \;\hfill\\
$^{4}$ Rochester Institute of Technology, NY, USA
\;\;\; \; \;\hfill\\
$^{5}$ Centre for Space Research, North-West University, Mahikeng 2745, South Africa
\;\;\; \; \;\hfill\\
$^{6}$ National Institute for Theoretical and Computational Sciences (NITheCS), South Africa
\\
\hfill\\
Correspondence:munalph@gmail.com\;\;\;\;\;\;\;\;\;\;\;\;\;\;\;\;\;\;\;\;\;\;\;\;\;\;\;\;\;\;\;\;\;\;\;\;\;\;\;\;\;\;\;\;\;\;\;\;\;\;\;\;\;\;\;\;\;\;\;\;\;\;\;\;\;\;\;\;\;\;\;\;\;\;\;
\begin{center}
\textbf{Abstract}
\end{center}
In this study, we present the evolution of cosmological perturbations in a universe
consisting of standard matter and interacting vacuum. We use the $1+3$ covariant formalism in
perturbation framework and consider two different models for the interacting vacuum; namely, a linear interacting model and interaction with creation pressure model. For both models, we derive the evolution equations governing the growth of linear perturbations for both radiation- and dust-dominated Universe. We find numerical solutions in appropriate limits, namely long and short wavelengths. For both models, the perturbations grow with time (decay with redshift), showing that structure formation is possible in an accelerated cosmic background. The perturbation amplitudes -- and their relative scalings with those of $\Lambda$CDM -- depend on the values of the interaction parameters considered, and in a way that can be used to constrain the models using existing and future large-scale structure data. In the vanishing limits of the coupling parameters of the interaction, we show that standard $\Lambda$CDM cosmology, both background and perturbed, is recovered.
\\
\hfill\\
\textit{keywords:} Cosmology--- dark matter--- dark energy--- interacting vacuum---cosmological constant--- covariant formalism--- Cosmological perturbations.This manuscript was accepted for publication in International Journal of Geometric Methods in Modern Physics.
\section{Introduction}\label{introduction}
Recent astronomical data show that the expansion of the Universe is accelerating \cite{riess1998observational,perlmutter2003supernovae}. Different hypotheses propose different mechanisms for such a cosmic acceleration, the most prominent proposal being a certain fluid with negative pressure known as dark energy. Among the widely explored candidates for dark energy is the cosmological constant $\Lambda$ \cite{lima2004thermodynamics, visinelli2019revisiting, tasitsiomi2003state}, which has now led to the development of the concordance cosmological model known as $\Lambda$CDM dominated by dark energy. The CDM part of $\Lambda$CDM represents [cold] dark matter, another sub-dominant component of the universe which is invisible in the electromagnetic spectrum but whose effects can be detected via gravitational interactions. The cosmological constant as the energy of free space (vacuum) suffers from the {\it cosmological constant problem} \cite{weinberg89, peebles2003cosmological, barrow2011value} and the {\it coincidence problem} \cite{velten14}, serious theoretical and observational issues that cannot and should not be ignored. Since the standard model of cosmology is based on General Relativity (GR), recent studies tried to propose different modified gravity theories \cite{debono2016general}.\\\\
\noindent One possible avenue to addressing the above issues related to the cosmological constant is treating $\Lambda$ as some limiting manifestation of  an interacting vacuum \cite{wands2012inhomogeneous, li2016testing, wang2014post, bruni2022nonsingular, mbonye2004cosmology}. For instance in the work done in \cite{bruni2022nonsingular}, it was found that one can produce nonsingular cosmology under the consideration of the interacting vacuum energy, and the cosmological constant appears as a late-time limiting case of the vacuum energy. In the work done by \cite{mbonye2004cosmology}, it was found that through the consideration of the creation pressure in the continuity equation led to the situation where the universe experiences quasi-periodic acceleration phases. On the other side, one finds a detailed exploration of interacting vacuum energy using Planck Data \cite{wang2014post}.\\\\
 Different formalism such as metric and the $1+3$ covariant formalisms can be used to study cosmological perturbations. The work done in 
\cite{geng2017matter} investigated the matter density perturbation and matter power spectrum in the running vacuum model using metric formalism. In addition to that, the work done in  \cite{perico2017running} studied linear scalar perturbations using metric formalism where they assumed the running vacuum as the sum of independent contributions associated with each of the matter species. On the other hand, the work done in  \cite{sharma2021growth} studied the growth of matter perturbations in an interacting dark energy scenario emerging from the metric-scalar-torsion couplings using metric formalism and obtained appropriate fitting formula for the growth index in terms of the coupling function and the matter density parameters. The  works done in \cite{borges2020growth, borges2008evolution, de2014inhomogeneous, yang2020forecasting} explored the perturbation aspects of interacting vacuum energy. In the work done by \cite{borges2008evolution, borges2008evolutionb}, the consideration of decaying vacuum was done focusing on the homogeneous interactions of both matter and dark energy. On the other hand the treatment of generalised Chaplygin gas models as inhomogeneous interacting dark energy with matter was done in \cite{de2014inhomogeneous} focusing on cosmic microwave background (CMB) anisotropies. In \cite{yang2020forecasting}, the authors investigated perturbations of interacting vacuum focusing on the  contributions of gravitational waves data, where they found a significant improvement in the CMB measurements. \\ \\   
The present paper aims to apply the $1+3$ covariant and gauge-invariant perturbations formalism to study large-scale structure formation scenarios for two interacting vacuum models treated in the works by Bruni et al \cite{bruni2022nonsingular} and Mbonye \cite{mbonye2004cosmology}. The $1+3$ covariant gauge-invariant formalism is used to study cosmological perturbations for both GR and modified gravity theories such as $f(R)$, $f(T)$ and $f(G)$. In the $1+3$ covariant formalism, the perturbation variables defined describe true physical degrees of freedom and no unphysical modes exist.\\\\
\noindent The rest of this paper is organised as follows: in Sec. \ref{back} we give a covariant description and the general linearised field equations involving the interacting dynamical vacuum. In Sec. \ref{pert} we define the covariant perturbation variables, derive their evolution equations and analyse their solutions. Finally in Sec. \ref{concsec} we discuss the results and give conclusions.\\\\
%
%%%%%%%%%%%%%%%%%%%%%%%%%%%%%%%%%%%%%%%%%%%%%%%%%%%%%%%%%%%%%%%%%%%%%%%%%%%%%%%%%%%%%%%%%%%%%%%%%%%
%
\noindent Natural units in which $c=8\pi G=1$
will be used throughout this paper, indices like a,b... run from $1$ to $3$ and Greek indices run from $0$ to $3$.
The symbols $\nabla$, $\D$ and the overdot $^{.}$ represent the usual covariant derivative, the spatial covariant derivative, and differentiation with respect to cosmic time, respectively. We use the
$(-+++)$ spacetime signature.  
\section{Background Field Equations}\label{back}
The standard GR gravitational action with a matter field contribution to the Lagrangian, ${\cal L}_m\,,$ is given by
\begin{equation}
 \mathcal{A}= \frac{1}{2} \int d^4x \sqrt{-g}\left[R+2{\mathcal{ L}_m} \right]\,.
\end{equation}
Using the variational principle of least action with respect to the metric $g_{ab}$,
the generalised Einstein Field Equations (EFEs) can be given in a compact form as
\begin{equation}
  G_{ab}= T_{ab}\,,
\end{equation}
 with the first (geometric) term represented by the Einstein tensor, and energy-momentum tensor of matter fluid forms given by 
\begin{equation}
 T_{ab} = \mu u_{a}u_{b} + ph_{ab}+ q_{a}u_{b}+ q_{b}u_{a}+\pi_{ab}\,,
\end{equation}
where $\mu$, $p$, $q_{a}$ and $\pi_{ab}$ are the energy density, isotropic pressure, heat flux and anisotropic pressure of the fluid, respectively. Here $u^{a}\equiv \frac{dx^{a}}{dt}$ is the $4$-velocity of fundamental observers comoving with the fluid. In a multi-component fluid universe filled with standard matter fields (dust, radiation, etc) and vacuum contributions, the total energy density, isotropic  pressures and heat flux are given, respectively, by $\rho=\rho_{m}+\rho_{v}$, $p=p_{m}+p_{v}$ and $q^{a}=q_{m}^{a}+q_{v}^{a}$, where $m$ and $v$ specify matter and vacuum respectively.
The vacuum equation of state parameter $w$ is given by $w=-1$. An arbitrary energy transfer $Q$ can reproduce  an arbitrary background cosmology with energy density \cite{wang2015reconstruction, wands2012inhomogeneous}
\begin{eqnarray}
 &&\rho=\rho_{m}+V\,,\\
 &&p=-V\,,
\end{eqnarray}
which reduces to $\varLambda CDM$ when $Q=0$ and we have a constant $V=\Lambda$.
The $4$-vector $Q_{\mu}$ can in general be decomposed into parallel and orthogonal parts to the $4$-velocity of the fluid
\begin{equation}
 Q^{a}=Q u^{a}+q^{a},
\end{equation}
where $q^{a}$ here is due to momentum exchange between matter and vacuum.
In this paper, we will assume an homogeneous isotropic model in which $Q^{a}$ is parallel to the matter $4$-velocity, $Q^{a}=Q u^{a}$. We will consider the case where the interactions reduce to pure energy exchange so that $q^{a}=0$ \cite{wang2014post, wang2013cosmological, borges2008evolution}.
Moreover for each interacting fluid, the following conservation equations considered in \cite{wands2012inhomogeneous, salvatelli2014indications, borges2020growth} hold:
\begin{eqnarray}
&& \dot{\rho}+3(1+\omega)H\rho=-Q\,,
 \label{eq7}\\
&& \dot{V}=Q\,.
 \label{eq8}
\end{eqnarray}
 The equation of state for the standard matter (such as dust and radiation) component is presented as $p=w \rho$, where $w$ is constant, and the total thermodynamic quantities $p=p_{m}+p_{v}$ and $\rho=\rho_{m}+\rho_{v}$ where the subscripts $m$ and $v$ stand for standard matter and vacuum contributions.
We define two different covariant choices for $Q$ as follows: the first one considered by Bruni et al \cite{bruni2022nonsingular}, hereafter referred to as Case 1, and the second by Mbonye  \cite{mbonye2004cosmology}, hereafter referred to as Case 2: 
\begin{eqnarray}
 &&Q_{1}=[\xi(V_{\varLambda}-V)+\sigma \rho]\theta\,,
 \label{eq9}\\
&& Q_{2}=\pi_{c}\theta\,,
 \label{eq10}
\end{eqnarray}
with $\pi_{c}=K[(3\gamma-2)\rho_{m}-2\rho_{v}]$. $\xi$, $\sigma$, $K$ and $\gamma$ are dimensionless coupling parameters and $V_{\varLambda}$ plays the role of an effective cosmological constant, and $\gamma=1+w$. Note that throughout the remainder of the paper, we refer to Case $1$ when using the interaction form $Q_{1}$ and Case $2$ when using $Q_{2}$.\\\\
Consider a spatially flat FRW universe with the metric
\begin{eqnarray}
 &&ds^{2}=-dt^{2}+a(t)^{2}(dx^{2}+dy^{2}+dz^{2}),
\end{eqnarray}
where the Friedman and the Raychaudhuri equations for flat spacetime are given respectively by
\begin{eqnarray}
 &&H^{2}\equiv \left(\frac{\dot{a}}{a}\right)^{2}=\frac{1}{3}(\rho+V)\,,
 \label{eq12}\\
 &&\dot{H}=-H^{2}-\frac{1}{6}\left[\left(1+3w\right)\rho-2V\right]\,,
 \label{eq13}
\end{eqnarray}
where $\theta=3H$, $H\equiv\frac{\dot{a}}{a}$ is the Hubble expansion rate, and $a(t)$ is the cosmological scale factor.
We dedicate the next subsections to analyse some of the
cosmological  solutions (in terms of the solutions for Hubble expansion and the deceleration parameters) to help us explain cosmic history of the late-time background due to the presence of the interacting vacuum, compared against the $\Lambda$CDM model.
\subsection{Background expansion for Case 1}
\begin{figure}
  \includegraphics[width=0.90\textwidth]{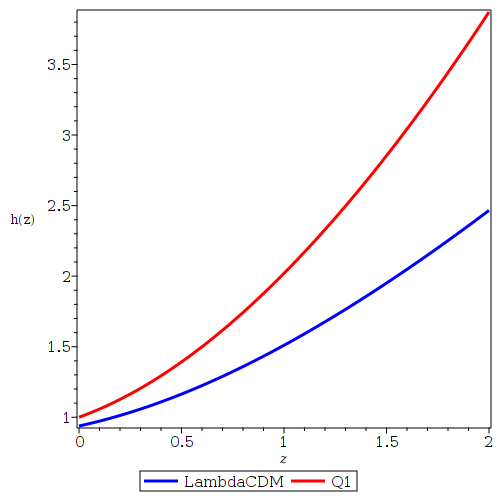}
  \caption{A plot of the normalised Hubble parameter, $h(z)$ versus redshift, $z$ for eq. \ref{eq17}.}
  \label{Fig1}
 \end{figure}

\begin{figure}
  \includegraphics[width=0.90\textwidth]{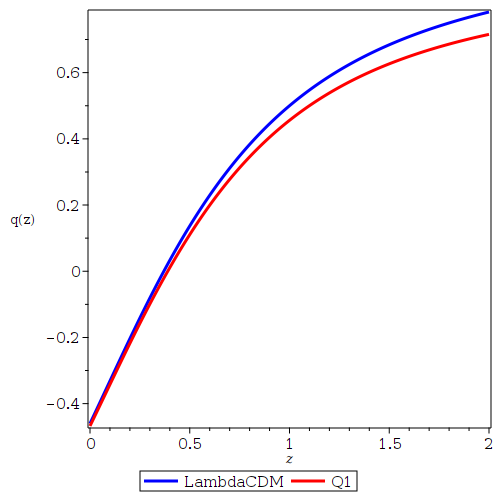}
  \caption{A plot of the deceleration parameter, $q(z)$ versus redshift, $z$ for eq. \ref{eq21}.}
  \label{Fig2}
 \end{figure}

We now consider linear models \cite{bruni2022nonsingular} presented by:
\begin{eqnarray}
 \rho&=&E_{1}a^{\alpha_{1}}+E_{2}a^{\alpha_{2}}\,,
 \label{eq14}\\
 V&=&V_{\Lambda}+\lambda_{1} a^{\alpha_{1}}+\lambda_{2}a^{\alpha_{2}}\,,
 \label{eq15}
\end{eqnarray}
where $E_{1}$, $E_{2}$, $\lambda_{1}$ and $\lambda_{2}$ are  energy contributions of the universe. Equation \ref{eq12} can be  represented as 
\begin{equation}
 H^{2}=\frac{1}{3}\left(E_{1}a^{\alpha_{1}}+E_{2}a^{\alpha_{2}}+V_{\Lambda}+\lambda_{1} a^{\alpha_{1}}+\lambda_{2}a^{\alpha_{2}}\right).
 \label{eq16}
\end{equation}
We can make change of variable and use redshift transformation as the redshift is a physically measurable quantity. Using $a=\frac{1}{1+z}$, here we assume a present value of scale factor to be $1$, eq. \ref{eq16} is transformed as 
\begin{equation}
 h_{1}(z)=\sqrt{(\Omega_{1}+\Omega_{3})(1+z)^{-\alpha_{1}}+ (\Omega_{2}+\Omega_{4})(1+z)^{-\alpha_{2}}+\Omega_{\Lambda}},
 \label{eq17}
\end{equation}
where $\Omega_{1}=\frac{E_{1}}{3H^{2}_{0}}$, $\Omega_{2}=\frac{E_{2}}{3H^{2}_{0}}$, $\Omega_{3}=\frac{\lambda_{1}}{3H^{2}_{0}}$, $\Omega_{4}=\frac{\lambda_{2}}{3H^{2}_{0}}$, $\Omega_{\Lambda}=\frac{V_{\Lambda}}{3H^{2}_{0}}$ and $h=\frac{H}{H_{0}}$.\\\\ 
The total fractional density parameter for a flat universe today is given by 
\begin{equation}
\Omega_{1}+\Omega_{2}+\Omega_{3}+\Omega_{4}+\Omega_{\Lambda}=1.
\end{equation}
The deceleration parameter $q$ is defined as
\begin{equation}
 q=-\frac{\ddot{a}a}{\dot{a}^{2}},
\end{equation}
which can be re-written (in redshift space) as
\begin{equation}
 q(z)=-1+(1+z)\frac{h'}{h}\,,
\end{equation}
where prime means derivative with respect to redshift $z$. The deceleration parameter in our case is presented as 
\begin{equation}
 q_{1}(z)=-\left(1+\frac{(\Omega_{1}+\Omega_{3})\alpha_{1}(1+z)^{-\alpha_{1}}+ (\Omega_{2}+\Omega_{4})\alpha_{2}(1+z)^{-\alpha_{2}}}{2h_{1}^{2}}\right).
 \label{eq21}
\end{equation}
Numerical results of normalised Hubble parameter (eq. \ref{eq17}) and deceleration parameter (eq. \ref{eq21}) are presented in Figs. \ref{Fig1} and \ref{Fig2}, respectively.
\subsection{Background expansion for Case 2}
\begin{figure}
  \includegraphics[width=0.90\textwidth]{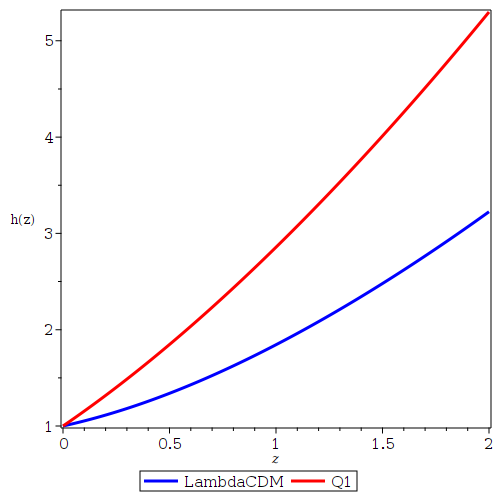}
  \caption{A plot of the normalised Hubble parameter, $h(z)$ versus redshift, $z$ for eq. \ref{eq24}.}
  \label{Fig3}
 \end{figure}

\begin{figure}
  \includegraphics[width=0.90\textwidth]{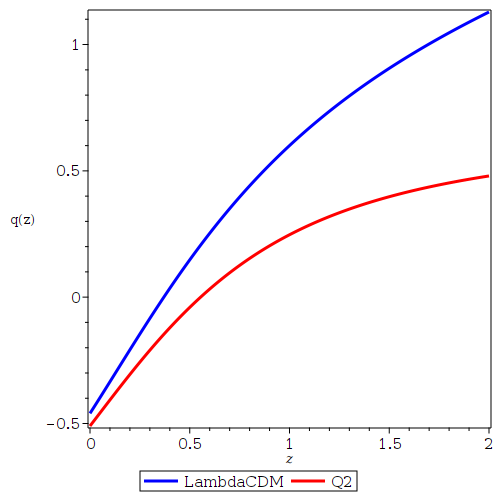}
  \caption{A plot of the deceleration parameter, $q(z)$ versus redshift, $z$ for eq. \ref{eq25}.}
  \label{Fig4}
 \end{figure}

 Considering $\rho_{m}=\rho_{0m}a^{3(1+\omega)}$, and $\rho_{v}=\rho_{0v}a^{\varSigma}$ \cite{mbonye2004cosmology} and using $a=\frac{1}{1+z}$, we get 
\begin{eqnarray}
 &&\rho_{m}=\rho_{0m}(1+z)^{-3(1+\omega)}\;,
 \label{eq22}\\
&& \rho_{v}=\rho_{0v}(1+z)^{-\varSigma},
 \label{eq23}
\end{eqnarray}
where $\varSigma=\beta \sigma(\psi)$, $\sigma=2+\sin(2\psi)$ \cite{mbonye2004cosmology} and $\beta$, $\rho_{0m}$ and $\rho_{0v}$ are coupling parameter, current values of the density of matter and vacuum, respectively.\\\\
Putting eq. \ref{eq22} and eq. \ref{eq23} into the Friedman equation (eq. \ref{eq12}) with some arrangements, the normalised Hubble parameter and deceleration parameter are given respectively as
 \begin{eqnarray}
&& h_{2}(z)=\sqrt{\Omega_{0d}(1+z)^{3}+\Omega_{0r}(1+z)^{4}+\Omega_{\Lambda}(1+z)^{\varSigma}}\;,
 \label{eq24}\\
&& q_{2}(z)=\left(-1+\frac{3\Omega_{0d}(1+z)^{3}+4\Omega_{0r}(1+z)^{4}+\varSigma \Omega_{\Lambda}(1+z)^{\varSigma}}{2h_{2}^{2}}\right),
 \label{eq25}
\end{eqnarray}
where $\Omega_{0d}=\frac{\rho_{0d}}{3H^{2}_{0}}$, $\Omega_{0r}=\frac{\rho_{0r}}{3H^{2}_{0}}$, $\Omega_{\Lambda}=\frac{\rho_{0v}}{3H^{2}_{0}}$, and 
%with 
\begin{equation}
 \Omega_{0d}+\Omega_{0r}+\Omega_{\Lambda}=1,
\end{equation}
is the total fractional density parameter for dust, radiation and cosmological constant respectively.\\\\
Numerical results of normalised Hubble parameter (eq. \ref{eq24}) and deceleration parameter (eq. \ref{eq25}) are presented in Fig. \ref{Fig3} and Fig. \ref{Fig4}, respectively. % 
To get numerical results, we considered $\alpha_{1}=-3$ and $\alpha_{2}=-4$ and note that for $\Lambda$CDM limit, $\Omega_{1}=\Omega_{0m}$, $\Omega_{2}=0=\Omega_{3}=\Omega_{4}$ and $\varSigma=0$  and the  deceleration parameter and the normalised Hubble parameter for the $\Lambda$CDM limit are given respectively as 
\begin{eqnarray}
 q(z)_{\Lambda CDM} &=& \left(\frac{\Omega_{0d}(1+z)^{3}+2\Omega_{0r}(1+z)^{4}-2\Omega_{0\Lambda}}{2h^{2}}\right)\,,\\ %~\\ \nonumber
 h(z)_{\Lambda CDM} &=& \sqrt{\Omega_{0d}(1+z)^{3}+\Omega_{0r}(1+z)^{4}+\Omega_{0\Lambda}}\,.
\end{eqnarray}
\noindent From the plots, in the reconstruction of $q(z)$, we have found that the evidence of the acceleration of the Universe happens at $z=0.39$ and $z=0.55$ for the first and second cases, respectively and for the $\Lambda$CDM limits, it happens at $z=0.38$ in agreement with the range $0.1\leq z\leq 0.6$ proposed in the work of \cite{gong2007reconstruction}. These are the transition redshifts when the universe underwent a phase change from deceleration to acceleration. It is found that at $z=0$, $h(z)=1$ for both cases. 
We will examine two different covariant choices of $Q$ in deriving perturbation equations \cite{wang2013cosmological, wang2014post} in the next sections.
\section{Perturbations and large-scale structure}\label{pert}
\subsection{General fluid description}
In this paper, we assume the interaction of matter fluid with vacuum energy in the universe and the growth of matter energy density fluctuations plays a significant role for structure formation.
We define the fluctuations in the energy overdensities of matter and the vacuum, as well as that of the expansion, in a covariant and gauge-invariant form, respectively, as follows \cite{carloni2006gauge, ananda2008detailed, abebe2015breaking}:
\begin{equation}
D^{m}_{a}=\frac{a\tilde{\bigtriangledown}_{a}\rho_{m}}{\rho_{m}}\;,\quad D^{v}_{a}=\frac{a\tilde{\bigtriangledown}_{a} V}{V}\;,\quad  Z_{a}=a\tilde{\bigtriangledown}_{a}\theta\,.
\label{eq29}
\end{equation}
All these quantities will be considered to develop the system of cosmological perturbations for interacting vacuum in the $1+3$ covariant formalism. 
\subsection{Linear evolution equations for the case  $Q=Q_{1}$}
In the case of eq. \ref{eq9} of linear interaction, eq. \ref{eq7} and eq. \ref{eq8} can be written as 
\begin{eqnarray}
 &&\dot{\rho}=-3H[(1+w+\sigma)\rho+\xi(V_{\varLambda-V})]\,,
 \label{eq32}\\
 &&\dot{V}=3H[\sigma \rho+\xi(V_{\varLambda}-V)]\,.
 \label{eq33}
\end{eqnarray}
These equations (eq. \ref{eq32} and eq. \ref{eq33}) form a coupled system of linear first order ordinary differential equations for $\rho$ and $V$. Using  scalar decomposition followed by harmonic decomposition and  redshift transformation technique as described in \textbf{Appendix B}, one can show that the $k$-th mode of the perturbations represented in redshift space evolves as:
 \begin{eqnarray}
 &&(1+z)h\Delta'_{m}+\Bigg\{\frac{3h}{1+w}\left[\left(1+w+\sigma\right)w+\left(1+2w\right)\xi\left( \frac{\Omega_{3}(1+z)^{-\alpha_{1}}+\Omega_{4}(1+z)^{-\alpha_{2}}}{\Omega_{1}(1+z)^{-\alpha_{1}}+\Omega_{2}(1+z)^{-\alpha_{2}}}\right)\right]\Bigg\}\Delta_{m}\nonumber\\
 &&\quad\quad-\Bigg\{\left(1+w+\sigma\right)
+\xi\left[ \frac{\Omega_{3}\left(1+z\right)^{-\alpha_{1}}+\Omega_{4}\left(1+z\right)^{-\alpha_{2}}}{\Omega_{1}\left(1+z\right)^{-\alpha_{1}}+\Omega_{2}\left(1+z\right)^{-\alpha_{2}}}\right] \Bigg\}\mathcal{Z}\nonumber\\
&&\quad\quad
+3\xi h\left[ \frac{\Omega_{\Lambda}+\Omega_{3}(1+z)^{-\alpha_{1}}+\Omega_{4}(1+z)^{-\alpha_{2}}}{\Omega_{1}(1+z)^{-\alpha_{1}}+\Omega_{2}(1+z)^{-\alpha_{2}}}\right]\Delta_{v}=0\,,
 \label{eq34}\\
 &&(1+z)h\mathcal{Z}'-\Bigg\{\frac{K^{2}}{1+w}\Bigg\{\left(1+3w\right)\frac{\Omega_{1}(1+z)^{-\alpha_{1}}+\Omega_{2}(1+z)^{-\alpha_{2}}}{2}\nonumber\\
 &&\quad\quad
 +w\left[\Omega_{\Lambda}+\Omega_{3}(1+z)^{-\alpha_{1}}+\Omega_{4}(1+z)^{-\alpha_{2}}\right]-w\left[3h^{2}+ \kappa^{2}(1+z)^{2}\right]\Bigg\}\Bigg\}\Delta_{m}-2h \mathcal{Z}\nonumber\\
 &&\quad\quad+K^{2}\left[\Omega_{\Lambda}+\Omega_{3}(1+z)^{-\alpha_{1}}+\Omega_{4}(1+z)^{-\alpha_{2}}\right]\Delta_{v}=0\,,
 \label{eq35}
 \end{eqnarray}
 \begin{eqnarray}
 &&(1+z)h\Delta'_{v}-3h \Bigg\{\xi+\sigma\left[\frac{\Omega_{1}(1+z)^{-\alpha_{1}}+\Omega_{2}(1+z)^{-\alpha_{2}}}{\Omega_{\Lambda}+\Omega_{3}(1+z)^{-\alpha_{1}}+\Omega_{4}(1+z)^{-\alpha_{2}}}\right] \nonumber\\
 &&\quad\quad+\xi\left[\frac{\Omega_{3}(1+z)^{-\alpha_{1}}+\Omega_{4}(1+z)^{-\alpha_{2}}}{\Omega_{\Lambda}+\Omega_{3}(1+z)^{-\alpha_{1}}+\Omega_{4}(1+z)^{-\alpha_{2}}}\right]\Bigg\}\Delta_{v}\nonumber\\ &&+\Bigg\{\frac{3h}{1+w}\Bigg\{\sigma\left[\frac{\Omega_{1}(1+z)^{-\alpha_{1}}+\Omega_{2}(1+z)^{-\alpha_{2}}}{\Omega_{\Lambda}+\Omega_{3}(1+z)^{-\alpha_{1}}+\Omega_{4}(1+z)^{-\alpha_{2}}}\right]\nonumber\\
 &&\quad\quad-w \xi\left[\frac{\Omega_{3}(1+z)^{-\alpha_{1}}+\Omega_{4}(1+z)^{-\alpha_{2}}}{\Omega_{\Lambda}+\Omega_{3}(1+z)^{-\alpha_{1}}+\Omega_{4}(1+z)^{-\alpha_{2}}}\right]\Bigg\}\Bigg\}\Delta_{m}\nonumber\\
 &&\quad\quad
 +\Bigg \{\sigma\left[\frac{\Omega_{1}(1+z)^{-\alpha_{1}}+\Omega_{2}(1+z)^{-\alpha_{2}}}{\Omega_{\Lambda}+\Omega_{3}(1+z)^{-\alpha_{1}}+\Omega_{4}(1+z)^{-\alpha_{2}}}\right]+\xi\left[\frac{\Omega_{3}(1+z)^{-\alpha_{1}}+\Omega_{4}(1+z)^{-\alpha_{2}}}{\Omega_{\Lambda}+\Omega_{3}(1+z)^{-\alpha_{1}}+\Omega_{4}(1+z)^{-\alpha_{2}}}\right]\Bigg\}\mathcal{Z}=0\,.
 \label{eq36}
\end{eqnarray}
where we have defined the dimensionless parameters
\begin{equation}
\frac{Z}{H_{0}}={\cal{Z}}\;,\quad\quad \kappa=\frac{k}{H_{0}}\,.
\end{equation}
For the $\varLambda$CDM limit, we set $\xi=\sigma=0$, and thus the system of equations (Eq. \ref{eq34} through to Eq. \ref{eq36}) reduce to:
\begin{eqnarray}
&& \Delta'_{m}+\frac{3w}{1+z} \Delta_{m}-\frac{1+w}{(1+z)h}\mathcal{Z}=0\,,
 \label{eq37}\\
&& \mathcal{Z}' -\frac{2}{1+z}\mathcal{Z}+\frac{w}{(1+w)(1+z)h}[3h^{2}+\kappa^{2}(1+z)^{2}]\Delta_{m}=0\;,
 \label{eq38}\\
&&  \Delta'_{v}=0\,.
  \label{eq39}
 \end{eqnarray}
In the following, we analyse the growth of energy density fluctuations from numerical results of Eq. \ref{eq34} through to Eq. \ref{eq39} for both  $\varLambda$CDM limits and perturbation equations in both the radiation- and dust-dominated epochs of the universe.
\subsection{Radiation-dominated epoch}
\begin{figure}[!htb]
\begin{minipage}{0.5\textwidth}
 %\centering
  \includegraphics[width=45mm]{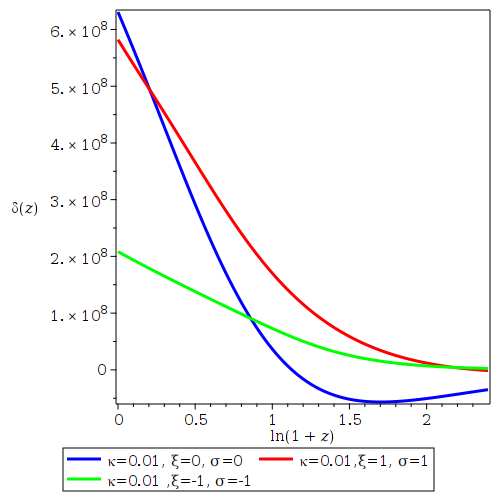}
  \caption{Plot of energy overdensity perturbation versus redshift in the radiation dominated universe in  long wavelength limits.}
  \label{Fig5}
 \end{minipage}
 \begin{minipage}{0.48\textwidth}
    %\centering
   \includegraphics[width=45mm]{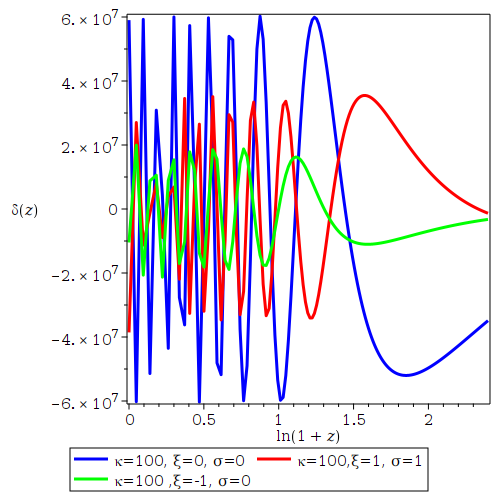}
  \caption{Plot of energy overdensity perturbation versus redshift in the radiation-dominated universe for  short wavelength limits}
  \label{Fig6}
 \end{minipage}
 \begin{minipage}{0.6\textwidth}
   \centering
   \includegraphics[width=45mm]{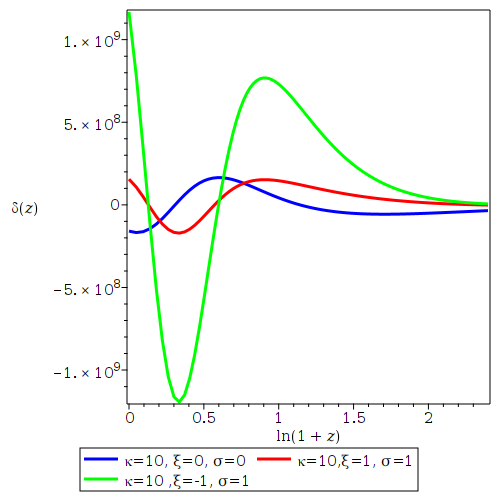}
  \caption{Plot of energy overdensity perturbation versus redshift in the radiation-dominated universe for short wavelength limits}
  \label{Fig7}
 \end{minipage}
 \end{figure}

For this epoch, we set $w=\frac{1}{3}$ and study the growth of the fluctuations \cite{dunsby1991gauge, dunsby1992covariant, ntahompagaze2017f} where Eqs. \ref{eq34} -- \ref{eq39} reduce to:
 \begin{eqnarray}
 &&(1+z)h\Delta'_{m}+\Bigg\{\frac{9h}{4}\left[\frac{1}{3}(\frac{4}{3}+\sigma)+\frac{5}{3}\xi\left[ \frac{\Omega_{3}(1+z)^{-\alpha_{1}}+\Omega_{4}(1+z)^{-\alpha_{2}}}{\Omega_{1}(1+z)^{-\alpha_{1}}+\Omega_{2}(1+z)^{-\alpha_{2}}}\right]\right]\Bigg\}\Delta_{m}\nonumber\\
 &&\quad\quad-\Bigg\{\left(\frac{4}{3}+\sigma\right)
+\xi\left[ \frac{\Omega_{3}(1+z)^{-\alpha_{1}}+\Omega_{4}(1+z)^{-\alpha_{2}}}{\Omega_{1}(1+z)^{-\alpha_{1}}+\Omega_{2}(1+z)^{-\alpha_{2}}}\right] \Bigg\}\mathcal{Z}\nonumber\\
&&\quad\quad+3\xi h\left[ \frac{\Omega_{\Lambda}+\Omega_{3}(1+z)^{-\alpha_{1}}+\Omega_{4}(1+z)^{-\alpha_{2}}}{\Omega_{1}(1+z)^{-\alpha_{1}}+\Omega_{2}(1+z)^{-\alpha_{2}}}\right]\Delta_{v}=0\,,
 \label{eq40}\\
 &&(1+z)h\mathcal{Z}'-\Bigg\{\frac{3K^{2}}{4}\Bigg\{\Omega_{1}(1+z)^{-\alpha_{1}}+\Omega_{2}(1+z)^{-\alpha_{2}}\nonumber\\
 &&\quad\quad+\frac{1}{3}\left[\Omega_{\Lambda}+\Omega_{3}(1+z)^{-\alpha_{1}}+\Omega_{4}(1+z)^{-\alpha_{2}}\right]-\frac{1}{3}\left[3h^{2}+ \kappa^{2}(1+z)^{2}\right]\Bigg\}\Bigg\}\Delta_{m}-2h \mathcal{Z}\nonumber\\
 &&\quad\quad+K^{2}\left[\Omega_{\Lambda}+\Omega_{3}(1+z)^{-\alpha_{1}}+\Omega_{4}(1+z)^{-\alpha_{2}}\right]\Delta_{v}=0\,,
 \label{eq41}\\
 &&(1+z)h\Delta'_{v}-3h \Bigg\{\xi+\sigma\left[\frac{\Omega_{1}(1+z)^{-\alpha_{1}}+\Omega_{2}(1+z)^{-\alpha_{2}}}{\Omega_{\Lambda}+\Omega_{3}(1+z)^{-\alpha_{1}}+\Omega_{4}(1+z)^{-\alpha_{2}}}\right]\nonumber \\
 &&\quad\quad+\xi\left[\frac{\Omega_{3}(1+z)^{-\alpha_{1}}+\Omega_{4}(1+z)^{-\alpha_{2}}}{\Omega_{\Lambda}+\Omega_{3}(1+z)^{-\alpha_{1}}+\Omega_{4}(1+z)^{-\alpha_{2}}}\right]\Bigg\}\Delta_{v}+\Bigg\{\frac{9h}{4}\Bigg\{\sigma\left[\frac{\Omega_{1}(1+z)^{-\alpha_{1}}+\Omega_{2}(1+z)^{-\alpha_{2}}}{\Omega_{\Lambda}+\Omega_{3}(1+z)^{-\alpha_{1}}+\Omega_{4}(1+z)^{-\alpha_{2}}}\right]\nonumber\\
 &&\quad\quad-\frac{1}{3} \xi\left[\frac{\Omega_{3}(1+z)^{-\alpha_{1}}+\Omega_{4}(1+z)^{-\alpha_{2}}}{\Omega_{\Lambda}+\Omega_{3}(1+z)^{-\alpha_{1}}+\Omega_{4}(1+z)^{-\alpha_{2}}}\right]\Bigg\}\Bigg\}\Delta_{m}\nonumber\\
 &&\quad\quad+\Bigg\{\sigma\left[\frac{\Omega_{1}(1+z)^{-\alpha_{1}}+\Omega_{2}(1+z)^{-\alpha_{2}}}{\Omega_{\Lambda}+\Omega_{3}(1+z)^{-\alpha_{1}}+\Omega_{4}(1+z)^{-\alpha_{2}}}\right]+\xi\left[\frac{\Omega_{3}(1+z)^{-\alpha_{1}}+\Omega_{4}(1+z)^{-\alpha_{2}}}{\Omega_{\Lambda}+\Omega_{3}(1+z)^{-\alpha_{1}}+\Omega_{4}(1+z)^{-\alpha_{2}}}\right]\Bigg\}\mathcal{Z}=0\,,
 \label{eq42}
 \end{eqnarray}
 with the corresponding $\Lambda$CDM limit given by
 \begin{eqnarray}
 &&\Delta'_{m}+\frac{1}{1+z} \Delta_{m}-\frac{4}{3(1+z)h}\mathcal{Z}=0\,,
 \label{eq43}\\
 &&\mathcal{Z}' -\frac{2}{1+z}\mathcal{Z}+\frac{1}{4(1+z)h}[3h^{2}+\kappa^{2}(1+z)^{2}]\Delta_{m}=0\,,
 \label{eq44}\\
  &&\Delta'_{v}=0\,.
  \label{eq45}
 \end{eqnarray}
We illustrate the background evolution (Eq. \ref{eq40} through to Eq. \ref{eq45}) as a function of redshift $z$ for different values of the parameters $\sigma$ and $\xi$ in Fig \ref{Fig5} for long wavelength modes and in Figs. \ref{Fig6} and \ref{Fig7} for short wavelength modes. In the particular case when $\sigma=0$ and $\xi=0$, the background evolution (Eq. \ref{eq43} through to Eq. \ref{eq45}) is identical to that for the $\Lambda$CDM model. We define the normalised energy density contrast for matter fluid as 
 $\delta(z)=\frac{\Delta^{k}_{m}(z)}{\Delta(z_{0})},$
where $\Delta(z_{0})$ is the matter energy density at the initial redshift.
%\pagebreak 
 \clearpage
\subsection{Dust-dominated epoch}
\begin{figure}[!htb]
\begin{minipage}{0.6\textwidth}
 %\centering
  \includegraphics[width=90mm]{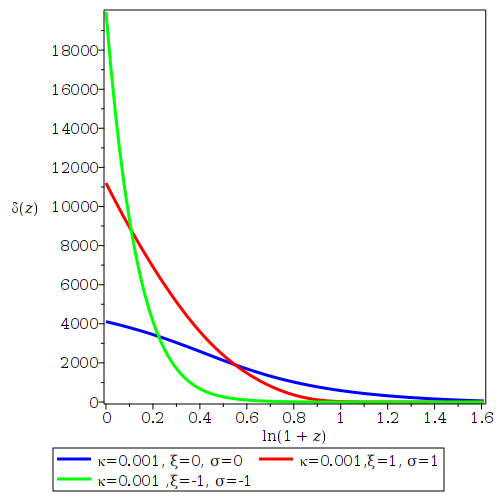}
  \caption{Plot of energy overdensity perturbation versus redshift in the dust-dominated universe for long wavelength limits.}
  \label{Fig8}
 \end{minipage}\\
 \begin{minipage}{0.6\textwidth}
    %\centering
   \includegraphics[width=90mm]{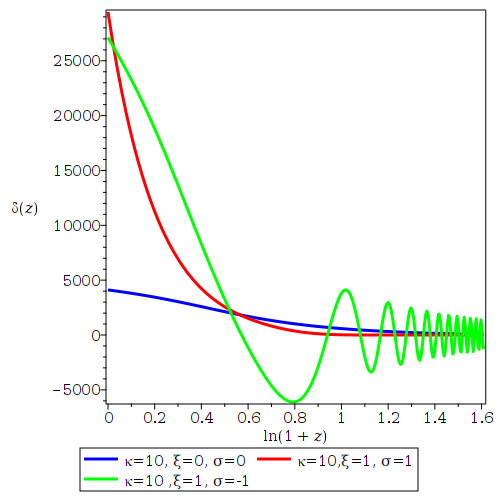}
  \caption{Plot of energy overdensity perturbation versus redshift in the dust-dominated universe for short wavelength limits.}
  \label{Fig9}
 \end{minipage}
  \end{figure}

We set $w=0$, for the dust-dominated epoch \cite{carloni2008evolution, ellis2009republication} so that Eq. \ref{eq34} through to Eq. \ref{eq39} reduce to
 \begin{eqnarray}
 &&(1+z)h\Delta'_{m}+\Bigg\{3h\xi\left[ \frac{\Omega_{3}(1+z)^{-\alpha_{1}}+\Omega_{4}(1+z)^{-\alpha_{2}}}{\Omega_{1}(1+z)^{-\alpha_{1}}+\Omega_{2}(1+z)^{-\alpha_{2}}}\right]\Bigg\}\Delta_{m} \nonumber \\ 
 &&\quad\quad-\Bigg\{1+\sigma
+\xi\left[\frac{\Omega_{3}(1+z)^{-\alpha_{1}}+\Omega_{4}(1+z)^{-\alpha_{2}}}{\Omega_{1}(1+z)^{-\alpha_{1}}+\Omega_{2}(1+z)^{-\alpha_{2}}}\right] \Bigg\}\mathcal{Z}\nonumber\\
&&\quad\quad+3\xi h\left( \frac{\Omega_{\Lambda}+\Omega_{3}(1+z)^{-\alpha_{1}}+\Omega_{4}(1+z)^{-\alpha_{2}}}{\Omega_{1}(1+z)^{-\alpha_{1}}+\Omega_{2}(1+z)^{-\alpha_{2}}}\right)\Delta_{v}=0\,,
 \label{eq46}\\
 &&(1+z)h\mathcal{Z}'-\left[K^{2}\frac{\Omega_{1}(1+z)^{-\alpha_{1}}+\Omega_{2}(1+z)^{-\alpha_{2}}}{2}\right]\Delta_{m}-2h \mathcal{Z}\nonumber\\
 &&\quad\quad+K^{2}\left[\Omega_{\Lambda}+\Omega_{3}(1+z)^{-\alpha_{1}}+\Omega_{4}(1+z)^{-\alpha_{2}}\right]\Delta_{v}=0\;,
 \label{eq47}\\
 &&(1+z)h\Delta'_{v}-3h \Bigg\{\xi+\sigma\left[\frac{\Omega_{1}(1+z)^{-\alpha_{1}}+\Omega_{2}(1+z)^{-\alpha_{2}}}{\Omega_{\Lambda}+\Omega_{3}(1+z)^{-\alpha_{1}}+\Omega_{4}(1+z)^{-\alpha_{2}}}\right] \nonumber\\
 &&\quad\quad+\xi\left[\frac{\Omega_{3}(1+z)^{-\alpha_{1}}+\Omega_{4}(1+z)^{-\alpha_{2}}}{\Omega_{\Lambda}+\Omega_{3}(1+z)^{-\alpha_{1}}+\Omega_{4}(1+z)^{-\alpha_{2}}}\right]\Bigg\}\Delta_{v}+\Bigg\{3h\sigma\frac{\Omega_{1}(1+z)^{-\alpha_{1}}+\Omega_{2}(1+z)^{-\alpha_{2}}}{\Omega_{\Lambda}+\Omega_{3}(1+z)^{-\alpha_{1}}+\Omega_{4}(1+z)^{-\alpha_{2}}}\Bigg\}\Delta_{m}\nonumber\\
 &&\quad\quad+\lbrace{\sigma\left[\frac{\Omega_{1}(1+z)^{-\alpha_{1}}+\Omega_{2}(1+z)^{-\alpha_{2}}}{\Omega_{\Lambda}+\Omega_{3}(1+z)^{-\alpha_{1}}+\Omega_{4}(1+z)^{-\alpha_{2}}}\right]+\xi\left[\frac{\Omega_{3}(1+z)^{-\alpha_{1}}+\Omega_{4}(1+z)^{-\alpha_{2}}}{\Omega_{\Lambda}+\Omega_{3}(1+z)^{-\alpha_{1}}+\Omega_{4}(1+z)^{-\alpha_{2}}}\right]\rbrace}\mathcal{Z}=0\,, %\nonumber\\
% &&
 \label{eq48}
 \end{eqnarray}
 with a corresponding $LCDM$ limit given by the set:
 \begin{eqnarray}
&& \Delta'_{m}-\frac{1}{(1+z)h}\mathcal{Z}=0\;,
 \label{eq49}\\
 &&\mathcal{Z}' -\frac{2}{1+z}\mathcal{Z}=0\;,
 \label{eq50}\\
  &&\Delta'_{v}=0\;.
  \label{eq51}
 \end{eqnarray}
Numerical results of eq. \ref{eq46} through to eq. \ref{eq51} are presented in Fig. \ref{Fig8} and Fig. \ref{Fig9} for long and short wavelength modes, respectively.
  \clearpage
\subsection{Linear evolution equations for the case $Q=Q_{2}$}
In the case \ref{eq10} of linear interaction, eq. \ref{eq7} and eq. \ref{eq8} can be written as:
\begin{eqnarray}
 &&\dot{\rho}=-K[(3\gamma-2)\rho_{m}-2\rho_{v}]\theta-\gamma\theta \rho_{m}\,,\\
 \label{eq52}
 &&\dot{V}=K[(3\gamma-2)\rho_{m}-2\rho_{v}]\theta-\frac{\varSigma}{3}\theta \rho_{v}.
 \label{eq53}
\end{eqnarray}
Using  scalar decomposition followed by harmonic decomposition  as described in \textbf{Appendix C} and the  redshift transformation technique, one can show that the $k$-th mode of the perturbations represented in redshift space evolves as: 
\begin{eqnarray}
 &&(1+z)h\Delta'_{m}+\Bigg\{\left[K(1+3w)w-2K(1+2w)\frac{\Omega_{\Lambda}}{\Omega_{0m}}(1+z)^{\varSigma+3(1+w)}\right]\frac{ 3h}{1+w}+3w h \Bigg\}\Delta_{m}\nonumber\\
 &&\quad\quad+\Bigg\{K(1+3w)+2K\frac{\Omega_{\Lambda}}{\Omega_{0m}}(1+z)^{\varSigma+3(1+w)}+(1+w)\Bigg\}\mathcal{Z}%\nonumber\\
 %&&\quad\quad
 +6Kh\frac{\Omega_{\Lambda}}{\Omega_{0m}}(1+z)^{\varSigma+3(1+w)}\Delta_{v}=0\;,
 \label{eq54}\\
 &&(1+z)h\mathcal{Z}'-2h \mathcal{Z}-\frac{1}{1+w}\Bigg\{\frac{K}{2}\left[ (1+3w)\Omega_{om}(1+z)^{-3(1+w)}+2w \Omega_{\Lambda}(1+z)^{\varSigma}\right]-3w h^{2}\nonumber\\
 &&\quad\quad-w \kappa^{2}(1+z)^{2}\Bigg\}\Delta_{m}+K\Omega_{\Lambda}(1+z)^{\varSigma}\Delta_{v}=0\;,
 \label{eq55}\\
 &&(1+z)h\Delta'_{v}+\Bigg\{\left[K(1+3w)\frac{\Omega_{0m}}{\Omega_{\Lambda}}(1+z)^{-3(1+w)-\varSigma}+2Kw +\frac{\varSigma}{3}w\right]\frac{3h}{1+w} \Bigg\}\Delta_{m}\nonumber\\
 &&\quad\quad+\Bigg\{K(1+3w)\frac{\Omega_{0m}}{\Omega_{\Lambda}}(1+z)^{-3(1+w)-\varSigma}-2K-\frac{\varSigma}{3}\Bigg\}\mathcal{Z}%\nonumber\\
 %&&\quad\quad
 -3K(3w+1)h\frac{\Omega_{0m}}{\Omega_{\Lambda}}(1+z)^{-3(1+w)-\varSigma}\Delta_{v}=0\,.\nonumber\\
 \label{eq56}
\end{eqnarray}
For the  $\Lambda$CDM limits, we set the coupling parameters
$K=0$ and $\beta=0$, so that we get 
\begin{eqnarray}
 &&\Delta'_{m}+\frac{3w}{1+z} \Delta_{m}-\frac{w+1}{(1+z)h}\mathcal{Z}=0\,,
 \label{eq57}\\
  &&\mathcal{Z}'-\frac{2}{1+z} \mathcal{Z}+\frac{w}{(1+w)(1+z)h}\left[3 h^{2}+\kappa^{2}(1+z)^{2}\right]\Delta_{m}=0\,,
 \label{eq58}\\
 &&\Delta'_{v}=0\,,
 \label{eq59}
\end{eqnarray}
where $h$ corresponds to $h_{2}$.
\subsection{Radiation-dominated epoch}
\begin{figure}[!htb]
\begin{minipage}{0.6\textwidth}
 %\centering
  \includegraphics[width=90mm]{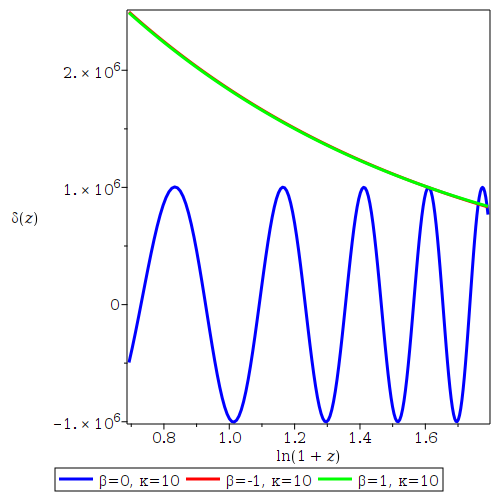}
   \caption{Plot of energy overdensity perturbation versus redshift in the radiation-dominated universe for short wavelength limits.}
  \label{Fig10}
 \end{minipage}\\
 \begin{minipage}{0.6\textwidth}
    %\centering
   \includegraphics[width=90mm]{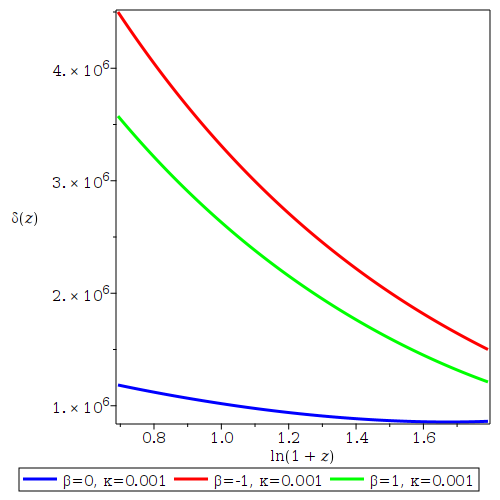}
  \caption{Plot of energy overdensity versus redshift in the radiation-dominated universe for long wavelength limits.}
  \label{Fig11}
 \end{minipage}
  \end{figure}

Setting $w=\frac{1}{3}$ in eq. \ref{eq54} through to eq. \ref{eq59}, the radiation-dominated perturbation equations can be represented as
\begin{eqnarray}
&& (1+z)h\Delta'_{m}+\Bigg\{\left(\frac{2}{3}K-\frac{10}{3}K\frac{\Omega_{\Lambda}}{\Omega_{0m}}(1+z)^{\varSigma+4}\right)\frac{ 9h}{4}+ h \Bigg\}\Delta_{m}\nonumber\\
&&\quad\quad-\Bigg\{2K-2K\frac{\Omega_{\Lambda}}{\Omega_{0m}}(1+z)^{\varSigma+4}+\frac{4}{3}\Bigg\}\mathcal{Z}+6Kh\frac{\Omega_{\Lambda}}{\Omega_{0m}}(1+z)^{\varSigma+4}\Delta_{v}=0\,,
 \label{eq60}\\
&& (1+z)h\mathcal{Z}'-2h \mathcal{Z}-\frac{3}{4}\Bigg\{K\left( \Omega_{om}(1+z)^{-4}+\frac{2}{3} \Omega_{\Lambda}(1+z)^{\varSigma}\right)-h^{2}\nonumber\\
&&\quad\quad-\frac{1}{3} \kappa^{2}(1+z)^{2}\Bigg\}\Delta_{m}+K\Omega_{\Lambda}(1+z)^{\varSigma}\Delta_{v}=0\,,
 \label{eq61}\\
&& (1+z)h\Delta'_{v}+\Bigg\{ \left[2K\frac{\Omega_{0m}}{\Omega_{\Lambda}}(1+z)^{-4-\varSigma}+\frac{2}{3}K +\frac{\varSigma}{9}\right]\frac{9h}{4} \Bigg\}\Delta_{m}\nonumber\\
&&\quad\quad+\Bigg\{2K\frac{\Omega_{0m}}{\Omega_{\Lambda}}(1+z)^{-4-\varSigma}-2K-\frac{\varSigma}{3}\Bigg\}\mathcal{Z}-6Kh\frac{\Omega_{0m}}{\Omega_{\Lambda}}(1+z)^{-4-\varSigma}\Delta^{k}_{v}=0\,,
 \label{eq62}
\end{eqnarray}
which further reduce to the following set of equations for $\Lambda$CDM:
\begin{eqnarray}
&& \Delta'_{m}+\frac{1}{1+z} \Delta_{m}-\frac{4}{3(1+z)h}\mathcal{Z}=0\,,
 \label{eq63}\\
&&  \mathcal{Z}'-\frac{2}{1+z} \mathcal{Z}+\frac{1}{4(1+z)h}[3 h^{2}+\kappa^{2}(1+z)^{2}]\Delta_{m}=0\;,
 \label{eq64}\\
&& \Delta'_{v}=0\,.
 \label{eq65}
\end{eqnarray}
\noindent Numerical results of eq. \ref{eq60} through to eq. \ref{eq65} are presented in Fig. \ref{Fig10} and Fig. \ref{Fig11} for short and long wavelength limits, respectively.
  \clearpage
\subsection{Dust-dominated epoch}
\begin{figure}
\begin{minipage}{0.6\textwidth}
 %\centering
   \includegraphics[width=90mm]{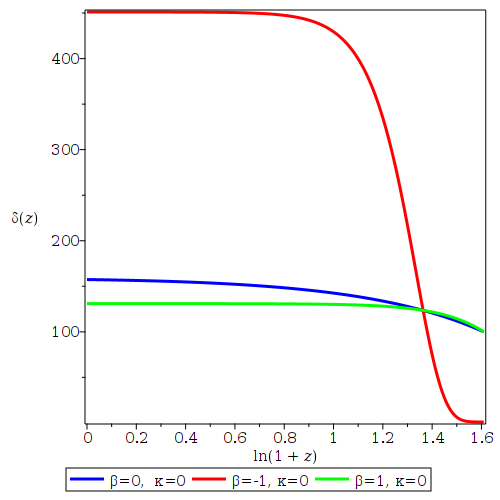}
  \caption{Plot of energy overdensity perturbation versus redshift in the dust-dominated universe for long wavelength limits.}
  \label{Fig12}
 \end{minipage}\\
 \begin{minipage}{0.5\textwidth}
    %\centering
   \includegraphics[width=90mm]{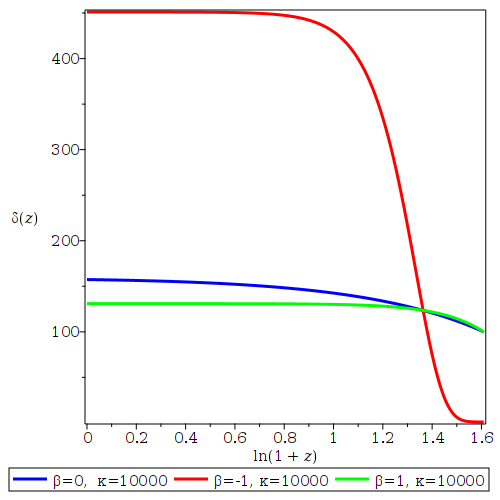}
  \caption{Plot of energy overdensity perturbation versus redshift in the dust-dominated universe for short wavelength limits.}
  \label{Fig13}
 \end{minipage}
  \end{figure}

Setting $w=0$ in eq. \ref{eq54} through to eq. \ref{eq59}, the dust-dominated perturbation equations can be presented as
\begin{eqnarray}
 &&(1+z)h\Delta'_{m}-6Kh\frac{\Omega_{\Lambda}}{\Omega_{0m}}(1+z)^{\varSigma+3}\Delta_{m}\nonumber\\
 &&\quad\quad+\Bigg\{K-2K\frac{\Omega_{\Lambda}}{\Omega_{0m}}(1+z)^{\varSigma+3}+1\Bigg\}\mathcal{Z}-6Kh\frac{\Omega_{\Lambda}}{\Omega_{0m}}(1+z)^{\varSigma+3}\Delta_{v}=0\,,
 \label{eq66}\\
 &&(1+z)h\mathcal{Z}'-2h \mathcal{Z}-\Bigg\{\frac{K}{2}\left[ \Omega_{om}(1+z)^{-3(1+\omega)}\right]\Bigg\}\Delta_{m}+K\Omega_{\Lambda}(1+z)^{\varSigma}\Delta_{v}=0\,,
 \label{eq67}\\
 &&(1+z)h\Delta'_{v}+3h\left[K\frac{\Omega_{0m}}{\Omega_{\Lambda}}(1+z)^{-3-\varSigma}\right] \Delta_{m}\nonumber\\ \nonumber
 &&\quad\quad+\Bigg\{K\frac{\Omega_{0m}}{\Omega_{\Lambda}}(1+z)^{-3-\varSigma}-2K-\frac{\varSigma}{3}\Bigg\}\mathcal{Z}-3Kh\frac{\Omega_{0m}}{\Omega_{\Lambda}}(1+z)^{-3-\varSigma}\Delta_{v}=0\,,
 \label{eq68}\\
 \end{eqnarray}
 with corresponding $\Lambda$CDM limits given by
 \begin{eqnarray}
 &&\Delta'_{m}-\frac{1}{(1+z)h}\mathcal{Z}=0\,,
 \label{eq69}\\
 && \mathcal{Z}'-\frac{2}{1+z} \mathcal{Z}=0\,,
 \label{eq70}\\
&& \Delta'_{v}=0.
 \label{eq71}
\end{eqnarray}
Numerical results of eq. \ref{eq66} through to eq. \ref{eq71} are presented in Fig. \ref{Fig12} and Fig. \ref{Fig13} for long and short wavelength limits, respectively.
  \clearpage
\section{Discussions and Conclusion}\label{concsec}
An interacting vacuum model provides an interesting alternative dark energy model which can be used to interpret the cosmological perturbations in the $1+3$ covariant formalism. Unlike other dark energy models such as scalar fields or modified gravity models, there are no additional degrees of freedom if the vacuum energy transfers energy-momentum to or from the considered matter fluids. In this paper, we have considered two different interacting vacuum models where the interaction is characterised by different dimensionless coupling parameters $K$, $\xi$, $\sigma$ and $\beta$ which produce the background dynamics of perturbation equations in the limit of vanishing interaction coupling parameters, where we recover the $\varLambda CDM$ cosmology. We have analysed  different cosmological parameters, namely the Hubble parameter and deceleration parameter for a flat universe. We presented the numerical results in  Fig. \ref{Fig1} through to Fig. \ref{Fig4} for both cases. From the plots, we see that $h(z)$ and $q(z)$  mimic the $\Lambda$CDM model. All results  are consistent with what is expected for the expansion history of the universe \cite{melchiorri2007did, hough2021confronting}.\\\\ 
We derived linear perturbation equations for both cases. Using different techniques such as scalar and harmonic decompositions, we get ordinary differential equations (ODEs). We transformed these ODEs into redshift equations and we get normalised perturbation equations in the redshift space. We further assume that the cases where the universe is filled with radiation and dust fluids and we considered the long wavelength modes where $\kappa\ll 1$ and the short wavelength modes within the horizon, where $\kappa \gg 1$ for both cases separately. We set $K=0=\sigma=\xi=\beta$ to recover the $\Lambda$CDM limit for both cases and $\beta=1$ to recover the case considered in the work done by Mbonye \cite{mbonye2004cosmology}.\\\\
For the first case considered in the work of Bruni et al \cite{bruni2022nonsingular}, the numerical results are presented in  Fig. \ref{Fig5} through to Fig. \ref{Fig7} for the radiation-dominated universe in both long and short wavelength modes respectively and Fig. \ref{Fig8} and Fig. \ref{Fig9} in dust-dominated universe for long and short wavelength modes respectively. We have explored different values of the coupling parameters to eventually notice the change in amplitude of the energy overdensities. For instance, the values $K,\sigma,\xi$  and $\beta$ vary from $-1$ to $1$ for both cases. \\ \\
We have analysed the growth of energy overdensity perturbations for both radiation and dust dominated universe for both cases. For radiation dominated , we set the equation of state parameter $w=\frac{1}{3}$ in Eq. \ref{eq32} through to Eq. \ref{eq38} to get equations  Eq. \ref{eq39} through to Eq. \ref{eq41}. We then illustrated the obtained perturbation equations for both long and short wavelength modes for different values of the coupling parameters. For long wavelength mode, the energy overdensity perturbations decay with increase in redshift. For short wavelength mode, the energy overdensity perturbations oscillate with decaying amplitude as the redshift increases.\\ In the dust dominated universe, we set $w=0$, in equations Eq. \ref{eq32} through to Eq. \ref{eq38} to get Eq. \ref{eq45} through to Eq. \ref{eq47}. We have then analysed  the long and short wavelength limits and find that the energy overdensity perturbations decay with increase in redshift for long wave modes, and decay then start to oscillate with decrease in amplitude  as the redshift increases. \\
For the $\Lambda CDM$ limits, we set $\sigma=\xi=0$ for both radiation and dust-dominated cases. 
From the plots, we notice that the energy overdensity perturbations decay with redshift for both radiation- and dust-dominated universe, long wavelength modes and show an oscillation behaviour for short wavelength modes.\\\\
For the second case considered in the work of Mbonye \cite{mbonye2004cosmology}, in a radiation-dominated universe, we set $w=\frac{1}{3}$ in Eq. \ref{eq53} through to Eq. \ref{eq58} to get  Eq. \ref{eq59} through to Eq. \ref{eq61} then we analysed  the energy overdensity for both long  and short wavelength modes and the numerical results are presented in  Fig. \ref{Fig10} and Fig. \ref{Fig11}. We see that for long wavelength modes, the energy density decays with increase in redshift for different values of coupling parameters. For short wavelength modes, the energy overdensity decays with increase in redshift for different values of $\beta$ and oscillates for $\beta=0$ as the redshift increases. For a dust-dominated universe, we set $w=0$ in Eq. \ref{eq53}  through to Eq. \ref{eq58} to get Eq. \ref{eq65} through to Eq. \ref{eq67}. the numerical results are presented in Fig. \ref{Fig12} and Fig. \ref{Fig13} for long and short wavelength modes, respectively. For long wavelength  and short wavelength limits the energy density overdensities decay with increase in redshift for different values of the coupling parameter $\beta$. The $\Lambda CDM$ case is recovered when one sets $K=\beta=0$ for both radiation and dust dominated universe.\\   From the plots, the energy overdensity decay with increasing redshift both for radiation- and dust-dominated universe. For both cases, all plots present the decay of energy overdensity with increase in redshift for both radiation- and dust-dominated epochs but there is an oscillation behaviour appearing in the first case which is not observed in the second case. The decay of energy overdensities with increase in redshift presents the possibility to explain large scale structure formation and cosmic acceleration scenarios.\\\\
In conclusion, both interacting-vacuum models support cosmic acceleration and the formation of large-scale structure, despite varying growth amplitudes that can help in constraining the parameters of the models vis-\`a-vis actual observed data -- and provide cosmological scenarios that are consistent with the $\Lambda$CDM model.
\section*{Acknowledgements}
AM acknowledges financial support from the Swedish International Development Agency (SIDA) through International Science Programme (ISP) to the  East African Astronomical Research Network (EAARN) (grant number AFRO:05). AA acknowledges that this work is based on the research supported in part by the National Research Foundation (NRF) of South Africa (grant number 112131). AM and AA (Abebe) acknowledge the hospitality of the Department of Physics of the University of Rwanda (UR), where this work was conceptualised. MM and JN acknowledge the financial support from SIDA through ISP to the UR through Rwanda Astrophysics Space and Climate Science Research Group (RASCRG), Grant number  RWA:01.
AA (Ayirwanda) and FT acknowledge UR-College of Science and Technology for the research facilities.
\appendix
\section{Useful Linearised Differential Identities}
For all scalars $f$, vectors $V_a$ and tensors that vanish in the background,
$S_{ab}=S_{\la ab\ra}$, the following linearised identities hold:
\begin{eqnarray}
\left(\D_{\la a}\D_{b\ra}f\right)^{.}&=&\D_{\la a}\D_{b\ra}\dot{f}-\sfrac{2}{3}\Theta\D_{\la a}\D_{b\ra}f+\dot{f}\D_{\la a}A_{b\ra}\label{a0}\;,\\
\ep^{abc}\D_b \D_cf &=& 0 \label{a1}\;, \\
\ep_{cda}\D^{c}\D_{\la b}\D^{d\ra}f&=&\ep_{cda}\D^{c}\D_{( b}\D^{d)}f=\ep_{cda}\D^{c}\D_{ b}\D^{d}f=0\label{a2}\;,\\
\D^2\left(\D_af\right) &=&\D_a\left(\D^2f\right) 
+\sfrac{1}{3}\tl{R}\D_a f \label{a19}\;,\\
\left(\D_af\right)^{\rd} &=& \D_a\dot{f}-\sfrac{1}{3}\Theta\D_af+\dot{f}A_a 
\label{a14}\;,\\
\left(\D_aS_{b\cdots}\right)^{\rd} &=& \D_a\dot{S}_{b\cdots}
-\sfrac{1}{3}\Theta\D_aS_{b\cdots}
\label{a15}\;,\\
\left(\D^2 f\right)^{\rd} &=& \D^2\dot{f}-\sfrac{2}{3}\Theta\D^2 f 
+\dot{f}\D^a A_a \label{a21}\;,\\
\D_{[a}\D_{b]}V_c &=& 
-\sfrac{1}{6}\tl{R}V_{[a}h_{b]c} \label{a16}\;,\\
\D_{[a}\D_{b]}S^{cd} &=& -\sfrac{1}{3}\tl{R}S_{[a}{}^{(c}h_{b]}{}^{d)} \label{a17}\;,\\
\D^a\left(\ep_{abc}\D^bV^c\right) &=& 0 \label{a20}\;,\\
\label{divcurl}\D_b\left(\ep^{cd\la a}\D_c S^{b\ra}_d\right) &=& {\ts{1\over2}}\ep^{abc}\D_b \left(\D_d S^d_c\right)\;,\\
\text{curlcurl} V_{a}&=&\D_{a}\left(\D^{b}V_{b}\right)-\D^{2}V_{a}+\sfrac{1}{3}\tl{R}V_{a}\label{curlcurla}\;,
\end{eqnarray}
\section{Perturbation equations for case $1$}
\subsection{Vector equations}
The first order linear evolution equations can be obtained by taking the temporal derivative of the gradient variables (eq. \ref{eq29}) and make use of equation \ref{a14} in the appendix which are presented as
\begin{eqnarray}
 &&\dot{D}_{a}^{m}-\Bigg\{\frac{\theta}{1+w}\left[(1+w+\sigma)w+(1+2w)\left(\frac{\xi }{\rho_{m}}(V_{\varLambda}-V)\right)\right]\Bigg\}D^{m}_{a}\nonumber\\
 &&\quad\quad+\left[\left(1+w+\sigma\right)
+\frac{\xi}{\rho_{m}} (V_{\varLambda}-V)\right]Z_{a}-\frac{\xi \theta}{\rho_{m}}VD_{a}^{v}=0\;,
 \label{eq72}\\
 &&\dot{Z}_{a}+\Bigg\{\frac{1}{1+w}\left[K^{2}\left((1+3w)\frac{\rho_{m}}{2}+w V\right)-\frac{\theta^{2}}{3}w\right]\Bigg\} D^{m}_{a}+\frac{w}{1+w}\tilde{\bigtriangledown}^{2}D^{m}_{a}+\frac{2}{3}\theta Z_{a}\nonumber\\
 &&\quad\quad-K^{2}VD^{v}_{a}=0\;,
 \label{eq73}\\
&& \dot{D}^{v}_{a}+\theta \left[\xi+\frac{\sigma \rho}{V}+\frac{\xi}{V}(V_{\Lambda}-V)\right]D^{v}_{a}-\Bigg\{\frac{\theta}{(1+w)V}\left[ \sigma \rho_{m}-w \xi\left(V_{\Lambda}-V\right)\right]\Bigg\}D^{m}_{a}\nonumber\\
&&\quad\quad-\frac{1}{V}\left(\sigma \rho+\xi(V_{\Lambda}-V)\right)Z_{a}=0\;. \label{eq74}
\end{eqnarray}
\subsection{Scalar decomposition}
Equation \ref{eq72} through to eq. \ref{eq74} are general perturbation equations. One needs to  extract the scalar part from these equations using scalar decomposition method \cite{abebe2012covariant, munyeshyaka2021cosmological}. The scalar parts of perturbation equations are believed to be responsible for spherical clustering of large scale structure.
The Scalar gradient variables are defined as 
\begin{equation}
\Delta_{m}=a\tilde{\nabla}^{a}D^{m}_{a},\;
 Z=a\tilde{\nabla}^{a}Z_{a}, \;
 \Delta_{v}=a\tilde{\nabla}^{a}D^{v}_{a}.
 \label{eq75}
\end{equation}
\subsubsection{Linear evolution equation for scalars}
The above scalar gradient variables evolve as presented below
\begin{eqnarray}
 &&\dot{\Delta}_{m}-\Bigg\{\frac{\theta}{1+w}\left[(1+w+\sigma)w+(1+2w)\left(\frac{\xi }{\rho_{m}}(V_{\varLambda}-V)\right)\right]\Bigg\}\Delta_{m}\nonumber\\
 &&\quad\quad+\left[\left(1+w+\sigma\right)
+\frac{\xi}{\rho_{m}} (V_{\varLambda}-V)\right]Z-\frac{\xi \theta}{\rho_{m}}V\Delta_{v}=0\;,
 \label{eq76}\\
 &&\dot{Z}+\left[\frac{1}{1+w}\left(K^{2}\left((1+3w)\frac{\rho_{m}}{2}+w V\right)-\frac{\theta^{2}}{3}w\right)\right]\Delta_{m}+\frac{w}{1+w}\tilde{\bigtriangledown}^{2}\Delta_{m}+\frac{2}{3}\theta Z\nonumber\\
 &&\quad\quad-K^{2}V\Delta_{v}=0\;,
 \label{eq77}\\
 &&\dot{\Delta}_{v}+\theta \left[\xi+\frac{\sigma \rho}{V}+\frac{\xi}{V}(V_{\Lambda}-V)\right]\Delta_{v}-\Bigg\{\frac{\theta}{(1+w)V}\left[\sigma \rho_{m}-w \xi(V_{\Lambda}-V)\right]\Bigg\}\Delta_{m}\nonumber\\
 &&\quad\quad-\frac{1}{V}\left(\sigma \rho+\xi(V_{\Lambda}-V)\right)Z=0\;. \label{eq78}
\end{eqnarray}
The scalar equations (Eq. \ref{eq76} through to Eq. \ref{eq78}) are used to analyse the energy density fluctuations by applying the harmonic decomposition method.
\subsection{Harmonic decomposition}
The standard harmonic decomposition of the evolution equations for perturbations can be used to transform first order linear differential equations into first order ordinary differential equations \cite{dunsby1992covariant, abebe2012covariant, munyeshyaka2021cosmological}. %
Define the eigeinfunctions of the covariant derivative with Laplace-Beltrami operator in a FRW spacetime as 
\begin{equation}
 \tilde{\nabla}^{2} M^{k}(x)=-\frac{k^{2}}{a^{2}}M^{k}(x),
\end{equation}
where $k$ is the wave-number related to the scale factor as $k=\frac{2\pi a}{\lambda}$, $\lambda$ is the wavelength of perturbations and $M^{k}(x)$ is the eigenfunctions of the covariant derivative.
Using the above decomposition scheme, the evolution equations (Eq. \ref{eq76} through to Eq. \ref{eq78}) are transformed as 
\begin{eqnarray}
 &&\dot{\Delta}^{k}_{m}-\Bigg\{\frac{\theta}{1+w}\left[(1+w+\sigma)w+(1+2w)\left(\frac{\xi }{\rho_{m}}(V_{\varLambda}-V)\right)\right]\Bigg\}\Delta^{k}_{m}\nonumber\\
 &&\quad\quad+\left[\left(1+w+\sigma\right)
+\frac{\xi}{\rho_{m}} (V_{\varLambda}-V)\right]Z^{k}-\frac{\xi \theta}{\rho_{m}}V\Delta^{k}_{v}=0\;,
 \label{eq80}\\
 &&\dot{Z}^{k}+\Bigg\{\frac{1}{1+w}\left[K^{2}\left((1+3w)\frac{\rho_{m}}{2}+w V\right)-w(\frac{\theta^{2}}{3}+ \frac{k^{2}}{a^{2}})\right]\Bigg\}\Delta^{k}_{m}+\frac{2}{3}\theta Z^{k}\nonumber\\
 &&\quad\quad-K^{2}V\Delta^{k}_{v}=0\;,
 \label{eq81}\\
 &&\dot{\Delta}^{k}_{v}+\theta \left[\xi+\frac{\sigma \rho}{V}+\frac{\xi}{V}(V_{\Lambda}-V)\right]\Delta^{k}_{v}-\Bigg\{\frac{\theta}{(1+w)V}\left[\sigma \rho_{m}-w \xi(V_{\Lambda}-V)\right]\Bigg\}\Delta^{k}_{m}\nonumber\\
 &&\quad\quad-\frac{1}{V}\left[\sigma \rho+\xi(V_{\Lambda}-V)\right]Z^{k}=0\;. \label{eq82}
\end{eqnarray}
\subsection{Redshift transformation}
The redshift transformation technique helps to transform equations into redshift space as the redshift is a physically measurable quantity which can  be used to compare cosmological behaviour  of the perturbation equations with cosmological observations \cite{sahlu2020scalar, ntahompagaze2020multifluid, sami2021covariant}. %
From
\begin{equation}
 a=\frac{1}{1+z},
 \label{eq83}
\end{equation}
and for convenience, we transform any time derivative functions $f$ into a redshift derivative as 
\begin{equation}
 \dot{f}=-(1+z)Hf',
 \label{eq84}
\end{equation}
where prime means derivative with respect to redshift ($\frac{\partial}{\partial z}$).
\section{Perturbation equations for case $2$}
\subsection{Vector equations}
Vector evolution equations for case $2$ are presented as
\begin{eqnarray}
 &&\dot{D}^{m}_{a}-\Bigg\{\left[K(1+3w)w-2K(1+2w)\frac{\rho_{v}}{\rho_{m}}\right]\frac{ \theta}{1+w}+w \theta \Bigg\}D^{m}_{a}\nonumber\\
 &&\quad\quad+\Bigg\{K(1+3w)-2K\frac{\rho_{v}}{\rho_{m}}+(1+w)\Bigg\}Z_{a}-2K\theta\frac{\rho_{v}}{\rho_{m}}D_{a}^{v}=0\;,
 \label{eq85}\\
 &&\dot{Z}_{a}+\frac{2}{3}\theta Z_{a}+\frac{1}{1+w}\left[\frac{K}{2}\left( (1+3w)\rho_{m}+2w\rho_{v}\right)-\frac{1}{3}w\theta^{2}\right]D^{m}_{a}+\frac{w}{w+1}\tilde{\bigtriangledown}^{2}D^{m}_{a}\nonumber\\
 &&\quad\quad-K\rho_{v}D^{v}_{a}=0\;,
 \label{eq86}\\
 &&\dot{D}^{v}_{a}-\Bigg\{\left[K(1+3w)\frac{\rho_{m}}{\rho_{v}}+2Kw +\frac{\varSigma}{3}wa\right]\frac{\theta}{1+w} \Bigg\}D^{m}_{a}-\Bigg\{K(1+3w)\frac{\rho_{m}}{\rho_{v}}-2K-\frac{\varSigma}{3}\Bigg\}Z_{a}\nonumber\\
 &&\quad\quad+K(3w+1)\theta\frac{\rho_{m}}{\rho_{v}}D_{a}^{v}=0\;.
 \label{eq87}
\end{eqnarray}
These linear evolution equations are general, we extract their scalar parts in the next subsection.
\subsection{Scalar equations}
Using the same transformation technique as in the first case, the scalar equations are presented as
\begin{eqnarray}
 &&\dot{\Delta}_{m}-\Bigg\{\left[K(1+3w)w-2K(1+2w)\frac{\rho_{v}}{\rho_{m}}\right]\frac{ \theta}{1+w}+w \theta \Bigg\}\Delta_{m}\nonumber\\
 &&\quad\quad+\Bigg\{K(1+3w)-2K\frac{\rho_{v}}{\rho_{m}}+(1+w)\Bigg\}Z-2K\theta\frac{\rho_{v}}{\rho_{m}}\Delta_{v}=0\;,
 \label{eq88}\\
 &&\dot{Z}+\frac{2}{3}\theta Z+\frac{1}{1+w}\Bigg\{\frac{K}{2}\left[ (1+3w)\rho_{m}+2w\rho_{v}\right]-\frac{1}{3}w\theta^{2}\Bigg\}\Delta_{m}+\frac{w}{w+1}\tilde{\bigtriangledown}^{2}\Delta_{m}\nonumber\\
 &&\quad\quad-K\rho_{v}\Delta_{v}=0\;,
 \label{eq89}\\
 &&\dot{\Delta}_{v}-\Bigg\{\left[K(1+3w)\frac{\rho_{m}}{\rho_{v}}+2Kw +\frac{\varSigma}{3}w\right]\frac{\theta}{1+w} \Bigg\}\Delta_{m}-\Bigg\{K(1+3w)\frac{\rho_{m}}{\rho_{v}}-2K-\frac{\varSigma}{3}\Bigg\}Z\nonumber\\
 &&\quad\quad+K(3w+1)\theta\frac{\rho_{m}}{\rho_{v}}\Delta_{v}=0\;.
 \label{eq90}
\end{eqnarray}
\subsection{Harmonic-decomposed perturbation equations}
The harmonically decomposed equations are given as follow:
\begin{eqnarray}
 &&\dot{\Delta}^{k}_{m}-\Bigg\{\left[K(1+3w)w-2K(1+2w)\frac{\rho_{v}}{\rho_{m}}\right]\frac{\theta}{1+w}+w \theta \Bigg\}\Delta^{k}_{m}\nonumber\\
 &&\quad\quad+\Bigg\{K(1+3w)-2K\frac{\rho_{v}}{\rho_{m}}+(1+w)\Bigg\}Z^{k}+2K\theta \frac{\rho_{v}}{\rho_{m}}\Delta^{k}_{v}=0\;,
 \label{eq91}\\
 &&\dot{Z}^{k}+\frac{2}{3}\theta Z^{k}+\frac{1}{1+w}\Bigg\{\frac{K}{2}\left[ (1+3w)\rho_{m}+2w\rho_{v}\right]-\frac{1}{3}w\theta^{2}-w \frac{k^{2}}{a^{2}}\Bigg\}\Delta^{k}_{m}\nonumber\\
 &&\quad\quad-K\rho_{v}\Delta^{k}_{v}=0\;,
 \label{eq92}\\
 &&\dot{\Delta}^{k}_{v}-\Bigg\{\left[K(1+3w)\frac{\rho_{m}}{\rho_{v}}+2Kw +\frac{\varSigma}{3}w\right]\frac{\theta}{1+w}\Bigg\}\Delta^{k}_{m}-\Bigg\{K(1+3w)\frac{\rho_{m}}{\rho_{v}}-2K-\frac{\varSigma}{3}\Bigg\}Z^{k}\nonumber\\
 &&\quad\quad+K(3w+1)\theta\frac{\rho_{m}}{\rho_{v}}\Delta^{k}_{v}=0\;.
 \label{eq93}
\end{eqnarray}

%\section*{Appendix}
 \noindent
{\color{blue} \rule{\linewidth}{1mm} }
  \end{document}